\documentclass[submission, Phys]{SciPost}
\usepackage{amsmath,amssymb,mathrsfs}
\usepackage{psfrag}
\usepackage{graphicx}
\usepackage{graphics}
\usepackage{epsfig}
\usepackage{bm}
\usepackage{color}
\usepackage{times}
\usepackage{verbatim,color,ulem}

\newcommand{\beq}{\begin{equation}}
\newcommand{\eeq}{\end{equation}}
\newcommand{\beqa}{\begin{eqnarray}}
\newcommand{\eeqa}{\end{eqnarray}}

\newcommand{\cblue}{\color{black}}

\begin{document}

\begin{center}{\Large \textbf
{Bose-Einstein condensate in an elliptical waveguide}
}\end{center}
\begin{center}
{Luca Salasnich}
\end{center}
\begin{center}
{Dipartimento di Fisica e Astronomia ``G. Galilei'' and QTech, 
Universit\`a di Padova, \\ via F. Marzolo 8, I-35131, Padova, Italy\\
Istituto Nazionale di Fisica Nucleare, Sezione di Padova, \\
Via Marzolo 8, 35131 Padova, Italy
\\
Istituto Nazionale di Ottica del Consiglio Nazionale delle Ricerche, 
Unita di Sesto Fiorentino, \\ 
Via Nello Carrara 2, 50019 Sesto Fiorentino, Italy}
\end{center}

\begin{center}
\today
\end{center}

\section*{Abstract}
{\bf We investigate the effects of spatial curvature for an atomic 
Bose-Einstein condensate confined in an elliptical waveguide. The 
system is well described by an effective 1D Gross-Pitaevskii equation 
with a quantum-curvature potential, which has the shape of a double-well 
but crucially depends on the eccentricity of the ellipse. 
The ground state of the system displays a quantum phase transition 
from a two-peak configuration to a one-peak configuration 
at a critical attractive interaction strength. In correspondence 
of this phase transition the superfluid fraction strongly reduces 
and goes to zero for a sufficiently attractive Bose-Bose interaction.\\}
\vspace{10pt}
\noindent\rule{\textwidth}{1pt}
\tableofcontents\thispagestyle{fancy}
\noindent\rule{\textwidth}{1pt}
\vspace{10pt}

\section{Introduction} 

How does a locally-varying spatial curvature 
influence the properties of low-dimensional quantum systems? 
This is a relevant question asked by 
scientists working in very different fields such as 
{\cblue the linear Schr\"odinger equation for a particle constrained 
on a curve manifold \cite{skinner,dacosta1,dacosta2}}, but also 
quantum gravity \cite{halliwell} or quantum chaos \cite{heller}. 
It is well know \cite{lee,jost,carmo} that the local curvature 
of a curve on the three-dimensional (3D) Euclidean 
space is characterized by the so-called geodesic curvature. 
This geodesic curvature $\kappa$ is an extrinsic quantity: 
it does not remain invariant if the curve is under the effect 
of a distance-preserving transformation \cite{lee,jost,carmo}. 
Instead, the local curvature of a surface on the 3D Euclidean space 
is characterized by the so-called Riemann curvature tensor, 
which can be written in terms of the invariant Gaussian 
curvature and the not-invariant average curvature \cite{lee,jost,carmo}. 
{\cblue The quantum motion of a particle on a curved waveguide 
has been anayzed by several authors \cite{exner1,jaffe,clark1,clark2,exner2,
exner-book}.} More recently, the highly nontrivial role of curvature 
for constrained quantum systems 
has been theoretically investigated with ultracold atomic gases 
confined in a quasi-1D \cite{leboeuf,schwartz,delcampo,sandin}
and quasi-2D configurations \cite{pelster}. 
The main result of {\cblue all these} investigations is that 
the local curvature gives rise to a quantum-curvature potential. 

In this paper we consider an atomic Bose-Einstein condensate (BEC) 
confined in a quasi-1D elliptical waveguide finding that the quantum-curvature 
potential has the shape of a double-well, if the eccentricity 
of the ellipse is different from zero. 
By numerically solving the 1D Gross-Pitaevskii equation of the 
BEC wavefunction under the effect of this quantum-curvature potential, 
we show that the ground state of the system is uniform along the waveguide 
only if the eccentricity $\epsilon$ of the ellipse is zero 
(circular waveguide with constant curvature). Instead, for $\epsilon\neq 0$ 
we find that the ground state is generically characterized by a two-peak 
configuration, where the peaks are located around the minima of the 
effective double-well potential. However, we discover that 
in the case of attractive interaction it exists a critical (negative) 
interaction strength below which the ground state exhibits 
a quantum phase transition from the two-peak configuration 
to a one-peak configuration. This is the analog of the spontaneous 
symmetry breaking, i.e. the modulational instability \cite{sala-prl},  
of the uniform configuration predicted some years 
ago for an 1D attractive BEC in a circular 
waveguide \cite{carr,kav,ueda}. Our results show that 
the critical interaction strength depends on the eccentricity $\epsilon$
of the ellipse in a non-trivial way. We also analyze 
the effect of a boost velocity on the BEC moving in the elliptical 
waveguide deriving the Leggett formula \cite{leggett} for the 
superfluid fraction of a 1D bosonic system \cite{rica,chomaz,martone}. 
Our numerical investigation reveals that the superfluid fraction 
decreases dramatically in response to this quantum phase transition, 
eventually reaching zero for a sufficiently negative Bose-Bose interaction. 

\section{Quantum-curvature potential} 

We consider a Bose-Einstein condensate (BEC) made of $N$ identical 
bosonic atoms of mass $m$. The atoms are constrained 
to move along a curve ${\cal C}$ 
by the presence of a strong harmonic potential 
of frequency $\omega_{\bot}$ 
in the local transverse plane with respect to ${\cal C}$. 
The characteristic length of the transverse confinement is 
$l_{\bot} = \sqrt{\hbar/(m\omega_{\bot})}$ 
where $\hbar$ is the reduced Planck constant. We introduce a 
local system $(s,u,v)$ of coordinates, where $s$ is the 
curvilinar abscissa (arclength) along ${\cal C}$ while 
$u$ and $v$ are two coordinates of the transverse plane 
\cite{leboeuf,schwartz,delcampo,sandin}. In this way 
the Lagrangian density of our problem is given by 
\beq
\mathscr{L} = {i\hbar\over 2} \left( \Psi^* \partial_t \Psi 
- \Psi \partial_t \Psi^* \right) - 
{\hbar^2\over 2m}|\nabla \Psi|^2 
- {m\omega_{\bot}^2\over 2} (u^2+v^2)|\Psi|^2 - {1\over 2} g |\Psi|^4 \; ,  
\eeq
where $\Psi(s,u,v,t)$ is the BEC wavefunction normalized to one 
and $g=4\pi \hbar^2a_s(N-1)/m$ is the 3D strength of the contact 
inter-atomic potential with $a_s$ the s-wave scattering length. 
Clearly, the Laplacian operator $\nabla^2$ must be written in 
terms of the local system  $(s,u,v)$ of 
coordinates \cite{leboeuf,schwartz,delcampo,sandin}. 
Assuming the factorization 
\beq 
\Psi(s,u,v,t) = \psi(s,t) \; {e^{-{(u^2+v^2)\over 2 \sigma(s,t)^2}} 
\over \pi^{1/2} \sigma(s,t)} 
\label{toosimple}
\eeq
and inserting this ansatz into the Lagrangian density, 
after integration over $u$ and $v$ one gets \cite{sandin,salas,toigo}, 
\beqa 
\bar{\mathscr{L}} &=& {i\hbar\over 2} 
\left( \psi^* \partial_t \psi - \psi \partial_t \psi^* \right) -  
{\hbar^2\over 2m}|\partial_s \psi|^2 + 
{\hbar^2 \kappa^2(s)\over 8m} |\psi|^2 
\nonumber 
\\
&-& \left( {\hbar^2\over 2m}{1\over \sigma^2} + 
{m\omega_{\bot}^2\over 2} \sigma^2 \right) |\psi|^2 
- {1\over 2} {g\over 2\pi\sigma^2} 
|\psi|^4  \; , 
\label{piripicchia}
\eeqa
where $\kappa(s)$ is the local geodesic curvature of ${\cal C}$, 
and the conditions $\sigma\kappa \ll 1$ and $\sigma\ll \xi$ 
must hold, with $\xi =\hbar/\sqrt{2g|\Psi|^2}$ 
the 3D healing length \cite{sandin}. 
The Euler-Lagrange equations of the 1D action functional 
with respect to the 1D wavefunction $\psi(s,t)$ 
and the transverse width $\sigma(s,t)$ are 
\beq
i\hbar \partial_t\psi = 
\Big[ -{\hbar^2\over 2m}\partial_s^2 
- {\hbar^2\kappa^2(s)\over 8m} + {\hbar^2\over 2m}{1\over \sigma^2} + 
{m\omega_{\bot}^2\over 2} \sigma^2 
+ {2\hbar^2a_s(N-1)\over m \sigma^2} |\psi|^2 \Big] \psi \; , 
\label{1dnpse}
\eeq
and 
\beq
\sigma^2 = l_{\bot}^2 \sqrt{1 + 2a_s (N-1) |\psi|^2}
\label{1dnpse-sigma}
\eeq
Eq. (\ref{1dnpse}), equipped with Eq. (\ref{1dnpse-sigma}), is 
the time-dependent 1D nonpolynomial Schr\"odinger equation (NPSE) 
\cite{salas,toigo} for the wavefunction $\psi(s,t)$ 
of the BEC moving along the curve ${\cal C}$ (see also \cite{sandin}). 
As previously discussed, the geodesic curvature $\kappa(s)$ gives 
rise to an effective potential 
\beq
U_{Q}(s) = - {\hbar^2\kappa(s)^2\over 8m} \; . 
\label{potQ}
\eeq
This curvature potential $U_Q(s)$ is quantum because 
it involves the square of the 
reduced Planck constant $\hbar$. At fixed atomic mass $m$, 
only if the square of the curvature $\kappa(s)$ is sufficently large 
the effects of this quantum-curvature potential become relevant. 

Under the assumption that $\sigma\simeq l_{\bot}$, 
which corresponds to a very strong transverse confinement,  
the 1D NPSE becomes the familiar 
1D Gross-Pitaevskii (GPE) equation
\beq
i\hbar \partial_t\psi = 
\Big[ -{\hbar^2\over 2m}\partial_s^2 
- {\hbar^2\kappa(s)^2\over 8m} + \hbar \omega_{\bot} 
+ {2\hbar^2a_s(N-1)\over m l_{\bot}^2} |\psi|^2 \Big] \psi \; .  
\label{td1dgpe}
\eeq
It is very important to stress that, from Eq. (\ref{1dnpse-sigma}), 
the condition $\sigma\simeq l_{\bot}$ implies $2a_s(N-1)|\psi|^2\ll 1$. 
In the rest of the paper we will work within this 1D regime. 
n the new version of the manuscript I shall discuss the role 
Clearly, Eq. (\ref{td1dgpe}) is reliable in the weak-coupling 
and strong-transverse-confinement regime, where both beyond-mean-field 
and transverse-size effects are very small.

\begin{figure}
\begin{center}
\includegraphics[width=12.cm,clip]{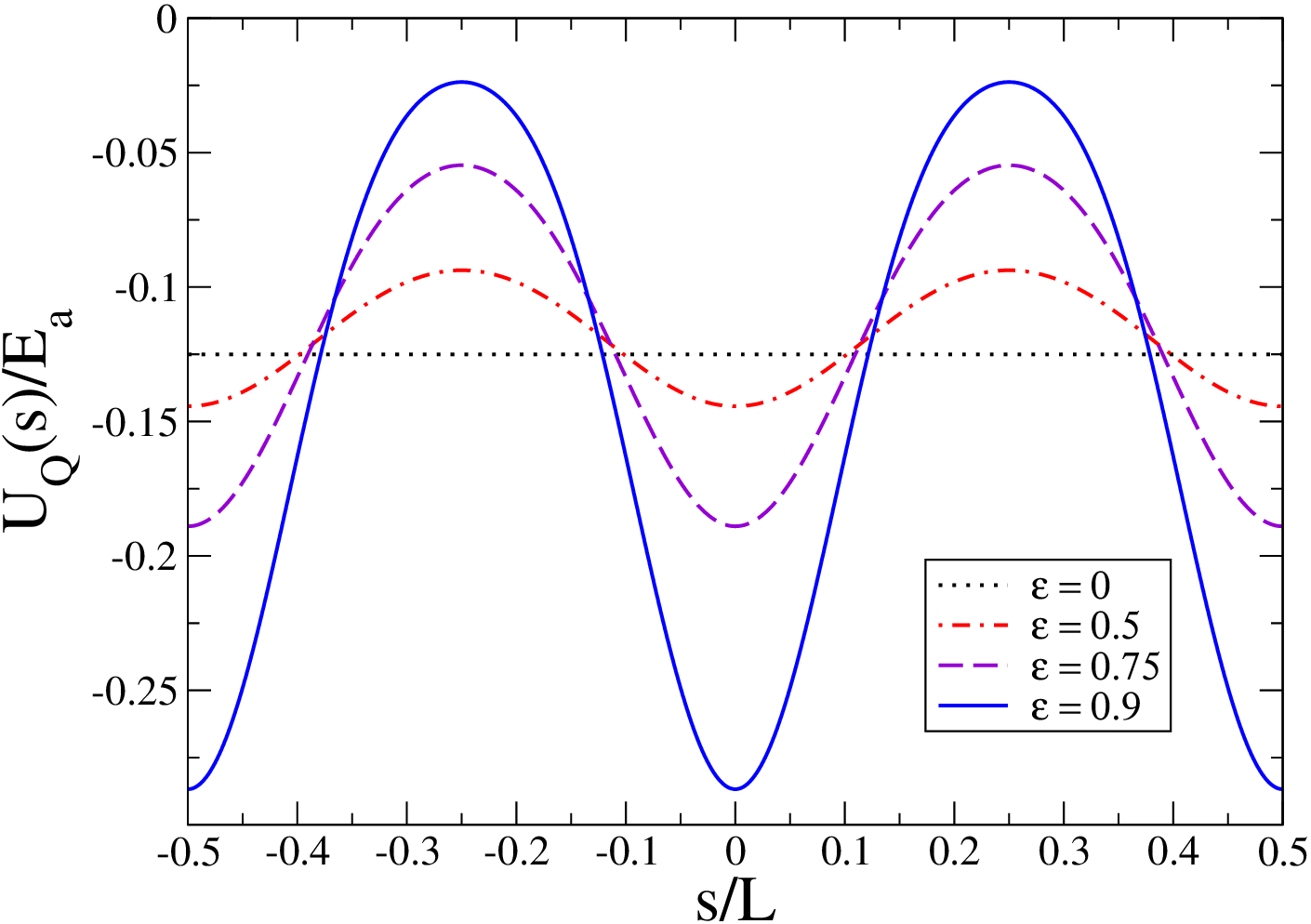}
\end{center}
\caption{Quantum-curvature potential 
$U_Q(s)$, Eq. (\ref{potQ}), induced by the geodesic curvature $\kappa(s)$
of an ellipse, as a function of the arclength $s$, where $a$ is the length 
of the major semi-axis, $L=aE(2\pi,\epsilon)$ is the perimeter of the ellipse, 
and $E_a=\hbar^2/(m a^2)$ a characteristic energy. 
The curves are obtained for different values of the eccentricity $\epsilon$.}
\label{fig1}
\end{figure}

\section{Properties of the elliptical waveguide}

We now choose an ellipse for the curve ${\cal C}$. By using cartesian 
coordinates its defining equation reads 
\beq 
{x^2\over a^2} + {y^2\over b^2} = 1 \; , 
\eeq
where $a$ and $b$ are the lengths of the two semi-axes of the ellipse. 
Here we assume that $a\geq b$, such that $a$ is the length of the 
major semi-axis. The eccentricity of the ellipse is defined as \cite{carmo}
\beq 
\epsilon = \sqrt{1-{b^2\over a^2}} \; . 
\eeq 
Clearly, $0\leq \epsilon < 1$ and for $\epsilon=0$ we obtain 
a circle of radius $R=a=b$. 
Introducing the angle $\phi \in [0,2\pi]$ we can write 
\beqa 
x = a \, \cos{(\phi)} 
\\
y = b \, \sin{(\phi)}
\eeqa
and the arclength $s$ along the ellipse can be expressed with 
the formula \cite{carmo} 
\beq 
s = a \, E(\phi , \epsilon) \; ,  
\label{s-ellipse}
\eeq
where 
\beq 
E(\phi,\epsilon) = \int_0^{\phi} 
\sqrt{1-\epsilon^2 \sin^2{(\phi')}} \, d\phi' 
\eeq
is the incomplete elliptic integral of the second kind. It follows 
that the perimeter $L$ of the ellipse reads 
\beq 
L = a \, E(2\pi ,\epsilon) \; ,  
\eeq
such that for $\epsilon=0$ we have $L=2\pi a$ because $E(2\pi,0)=2\pi$. 
Instead, for $\epsilon \to 1$ we have $L\to 4a$ because $E(2\pi,1)= 4$. 
We conclude that $4a<L\leq 2\pi a$. 
The geodesic curvature $\kappa$ of the ellipse can be written as \cite{carmo}
\beq 
\kappa = {1\over a} \; {\sqrt{1-\epsilon^2} \over 
\left( \sin^2{(\phi)} + \sqrt{1-\epsilon^2} 
\, \cos^2{(\phi)} \right)^{3/2} } \; . 
\label{k-ellipse}
\eeq
At fixed $a$ and $\epsilon$, the maximum of the curvature 
is obtained for $\phi=0$ and $\phi=\pi$, i.e. 
\beq 
\kappa_{max} =  {1\over a} \; {1 \over 
\left(1-\epsilon^2 \right)^{1/4} }  \; , 
\eeq
while the minimum of the curvature is obtained 
for $\phi=\pi/2$ and $\phi=3\pi/2$, i.e.
\beq  
\kappa_{min} = {1\over a} \; \sqrt{1-\epsilon^2}  \; . 
\eeq
Thus, for $\epsilon=0$ we have $\kappa_{max}=\kappa_{min}=1/a$ 
while for $\epsilon\to 1$ we have $\kappa_{max}\to +\infty$ 
and $\kappa_{min}\to 0$. We conclude that $1/a\leq k_{max}<+\infty$ 
and $0< k_{min}\leq 1/a$. 
In general, the formula which gives the curvature 
$\kappa$ as a function of the arclength $s$ is called Cesaro 
equation. Unfortunately, in the case of the ellipse 
there is no Cesaro equation. In other words, an analytical 
formula of $\kappa$ as a function of $s$ is not available. 
However, from Eqs. (\ref{s-ellipse}) and (\ref{k-ellipse}), 
fixing the length $a$ and the eccentricity $\epsilon$ 
of the ellipse, we can easily plot $\kappa$ vs $s$ using $\phi$ 
as dummy variable. More explicitly: we calculate separately 
$\kappa$ vs $\phi$ and $s$ vs $\phi$, and then we plot $\kappa$ vs $s$.  
The curvature $\kappa(s)$ has a the periodic structure of $\kappa(s)$. 
By increasing $\epsilon$, the perimeter $L$ of the ellipse 
slightly decreases while $\kappa_{max}$ and $\kappa_{min}$ pull away. 
In Fig. \ref{fig1} we plot the quantum-curvature 
potential $U_Q(s)$, Eq. (\ref{potQ}), 
induced by the curvature $\kappa(s)$ of the ellipse for 
different values of the eccentricity $\epsilon$. 
The figure clearly shows 
that,  for $\epsilon\neq 0$, $U_Q(s)$ is symmetric double-well potential 
where the depth of the wells becomes larger by increasing the eccentricity. 
The minima (maxima) of the quantum-curvature potential $U_Q(s)$ 
are in correspondence to the maxima (minima) of the curvature $\kappa(s)$. 

\begin{figure}
\begin{center}
\includegraphics[width=12.cm,clip]{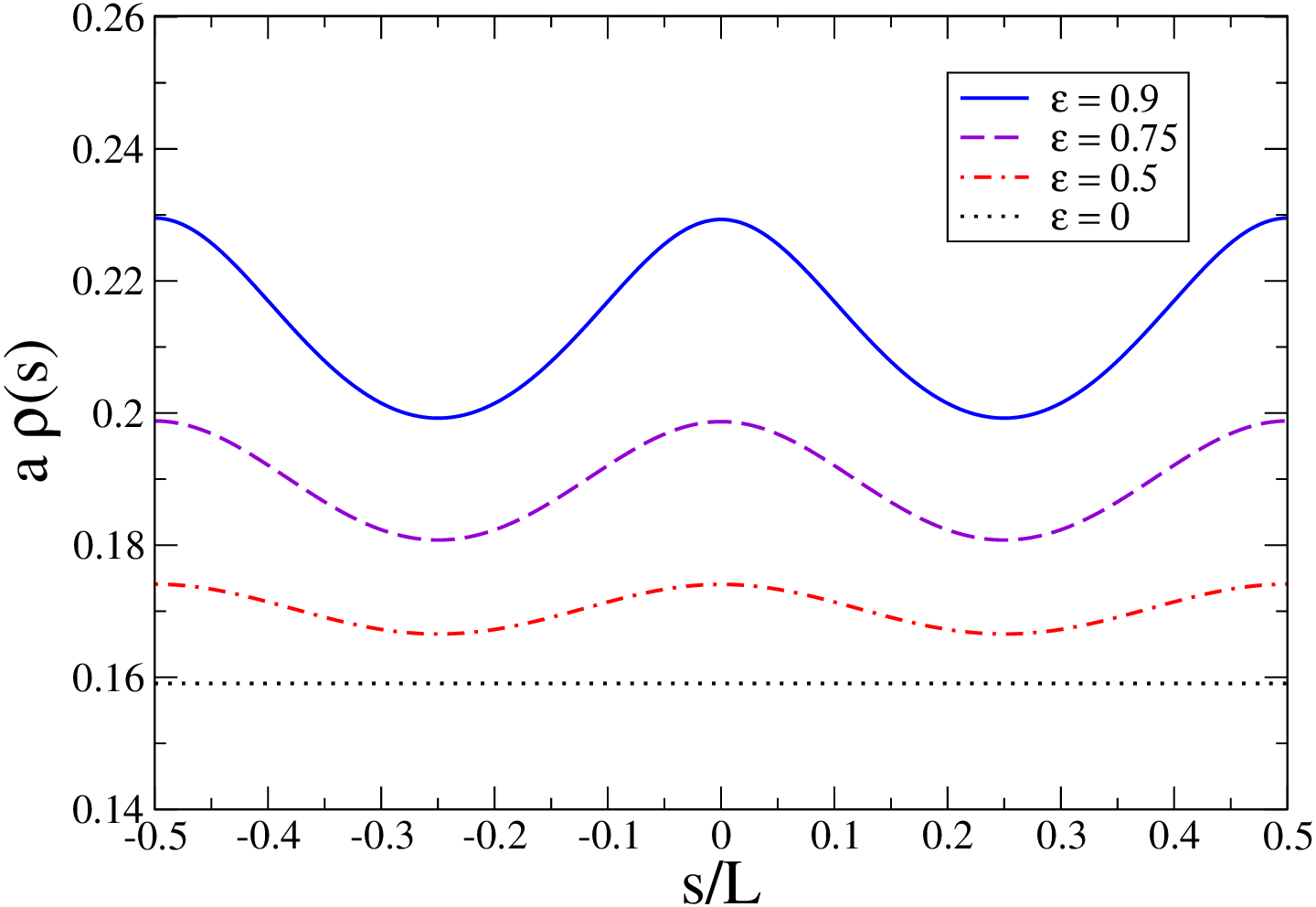}
\end{center}
\caption{Probablility density $\rho(s)$ of the non-interacting BEC ground state 
in an ellipse as a function of the arclength $s$, where $a$ is the length 
of the major semi-axis and $L=aE(2\pi,\epsilon)$ is the perimeter of the 
ellipse. The curves are obtained for different 
values of the eccentricity $\epsilon$. Here the s-wave scattering 
length $a_s$ is set to zero or, equivalently, the number $N$ 
of particles is set to one.} 
\label{fig2}
\end{figure}

\section{BEC ground state in elliptical waveguide} 

The time-independent 1D GPE is obtained from Eqs. (\ref{td1dgpe}) setting 
\beq 
\psi(s,t) = \Phi(s) \, e^{-i (\mu+\hbar\omega_{\bot}) t/\hbar} \; .  
\eeq
In this way we have 
\beqa 
\mu \, \Phi = 
\Big[ -{\hbar^2\over 2m}\partial_s^2 
- {\hbar^2\kappa(s)^2\over 8m} + 
{2a_s(N-1)\hbar^2 \over m l_{\bot}^2} |\Phi|^2 \Big] \Phi \; , 
\label{1dgpe}
\eeqa
that is the 1D GPE equation for the stationary wavefunction $\Phi(s)$, 
such that $\rho(s) = |\Phi(s)|^2$ 
is the probability density of finding the BEC at the position $s$. 
In Fig. \ref{fig2} we report the probability density $\rho(s)$ 
of the ground-state of the non-interacting ($a_s=0$ or, 
equivalently, $N=1$) BEC confined along the 
ellipse as a function of the arclength $s$. Our results are obtained by solving 
Eq. (\ref{td1dgpe}) with a Crank-Nicolson predictor-corrector method 
and imaginary time. In the figure, the curves correspond to 
different values of the eccentricity $\epsilon$. Clearly, for $\epsilon=0$ 
the ground state is uniform along the ellipse. However, 
for $\epsilon\neq 0$ the ground state is no more uniform due to a 
non-constant curvature $\kappa(s)$ which implies a non-constant 
effective potential $U_Q(s)=-\hbar^2\kappa(s)^2/(8m)$. By increasing 
the eccentricity $\epsilon$ the localization of $\rho(s)$ 
around the minima of $U_Q(s)$, where the curvature is larger, 
becomes more evident. 

\begin{figure}
\begin{center}
\includegraphics[width=12.cm,clip]{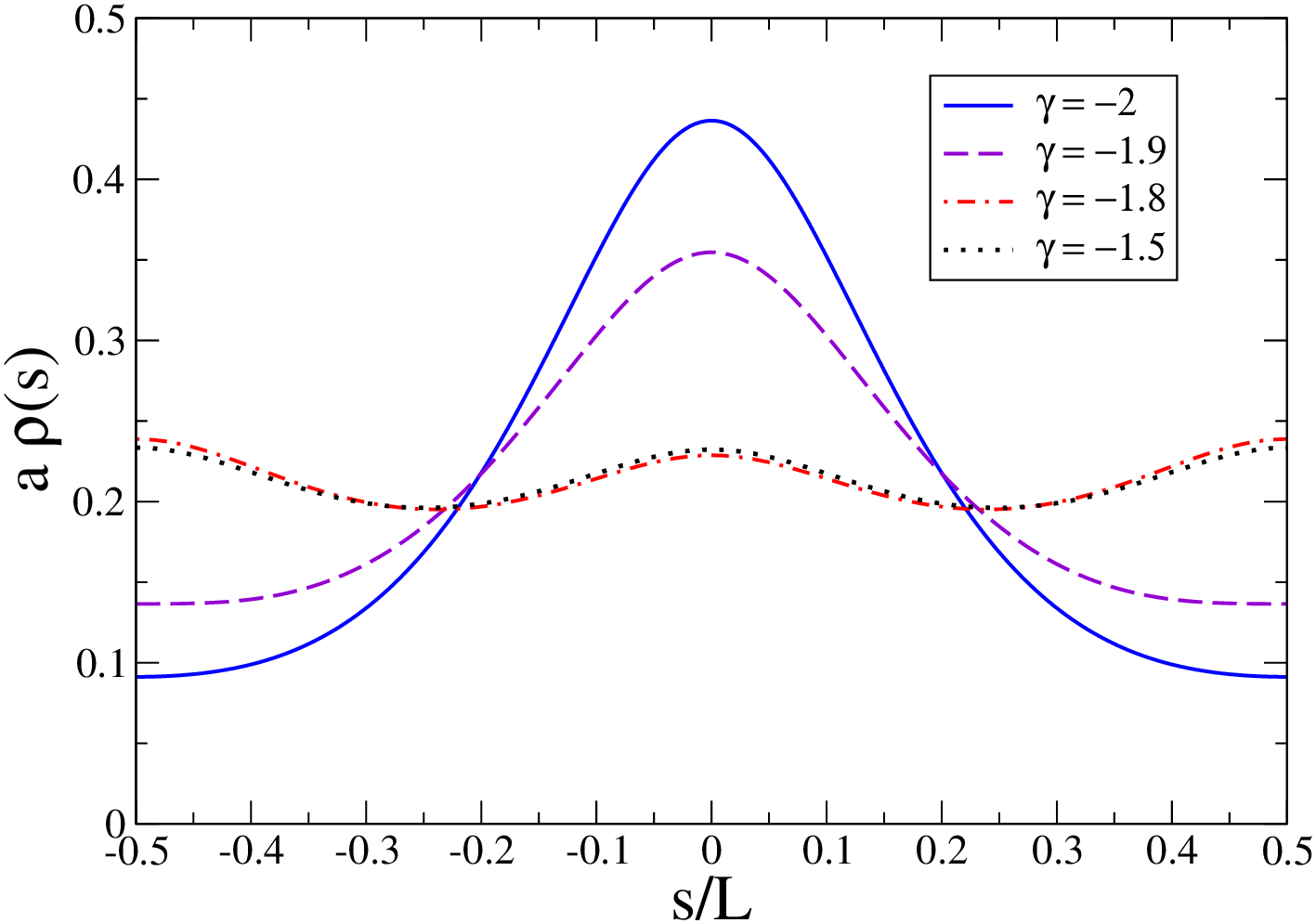}
\end{center}
\caption{Probablility density $\rho(s)$ of the attractive BEC ground state 
in an ellipse as a function of the arclength $s$, where $a$ is the length 
of the major semi-axis and $L=aE(2\pi,\epsilon)$ is the perimeter of the 
ellipse. The curves are obtained with eccentricity $\epsilon=0.9$ 
for different values of the adimensional 
interaction strength $\gamma=2aa_s(N-1)/l_{\bot}^2$, where $a_s$ is the 3D 
s-wave scattering length and $l_{\bot}$ is the characteristic length of the 
transverse harmonic confinement.}
\label{fig3}
\end{figure}

\section{Quantum phase transition} 

It is interesting to investigate the effect of the inter-atomic 
interaction on the ground state properties of the system. 
In adimensional units the interaction strength reads 
$\gamma = g/(2\pi l_{\bot}^2 aE_a)=2 a a_s(N-1)/l_{\bot}^2$ 
with $E_a=\hbar^2/(m a^2)$. 
In Fig. \ref{fig3} we consider the case of an attractive BEC 
and plot our numerical results obtained 
for different values of a negative $\gamma$ with 
fixed eccentricity $\epsilon=0.9$. Quite remarkably, 
for $\gamma<-1.5$ the ground state has a spontaneous symmetry breaking: 
one of the two local mimima contains more bosons. Indeed for 
$\gamma=-2$ this single-well localization is very clear. 
It is important to observe that this kind of quantum phase transition happens 
for any $\epsilon$. In the case of a circle ($\epsilon=0$) a similar 
spontaneous symmetry breaking was predicted about 
20 years ago \cite{carr,ueda,kav}. Actually, this 
quantum phase transition, or spontaneous symmetry breaking, 
is nothing else than the modulational instability of the ground-state 
configuration, induced by the appearance of an imaginary component 
in the energies of the elementary excitations of the 
ground state \cite{sala-prl}. However, for $\epsilon=0$ there 
is a quantum phase transition from a uniform configuration to a single-peak 
configuration, while for the $\epsilon\neq 0$ there is a quantum phase 
transition from a two-peak configuration to a single-peak configuration. 
This quantum phase transition was observed some years ago 
with an attractive BEC of $^{39}$K atoms, where the double-well potential 
was created by intersecting two pairs of laser beams \cite{fattori}. 
Our double-well system is slightly different because the particles tunnel 
from one well to the other well following two different curved paths; 
moreover, our elipsoidal configuration offers also the possibility 
of having persistent currents. 
For a sufficiently attractive BEC the transverse width $\sigma$ 
of the BEC becomes smaller than $l_{\bot}$ and there is the collapse of the 
single-peak configuration. For $\epsilon=0$ the 
1D NPSE predicts the collapse of this single-peak configuration 
at $\gamma_c=(4/3)(a/l_{\bot})$ \cite{salas,parola}. Thus, 
our numerical results of Fig. \ref{fig3}, obtained from the 1D GPE, 
are fully reliable under the condition $a/l_{\bot}\gg 1$. 
This is again the condition of a tight transverse confinement. 

\section{Superfluid fraction} 

Let us consider the effect of a boost velocity $v_{B}$ on the BEC 
moving along the ellipse. 
In this case Eq. (\ref{1dgpe}) is modified as follows 
\beq
\mu \, \Phi = 
\Big[ {1\over 2m} \left(-i\hbar \partial_s-
mv_{\mathrm{B}}\right)^{2} - {\hbar^2\kappa^2(s)\over 8m} 
+ {2a_s(N-1)\hbar^2 \over m l_{\bot}^2} |\Phi|^2 \Big] \Phi \; , 
\label{s1dh}
\eeq
Now we set 
\beq 
\Phi(s)= {n(s)^{1/2}\over \sqrt{N}} \, e^{i \theta(s)} \; , 
\label{standard}
\eeq 
where $n(s)=N\rho(s)$ is the local number density of the BEC. Morover, 
we introduce the local velocity field 
\beq 
v(s) = {\hbar\over m}\partial_s \theta(s) \; . 
\label{vcrucial}
\eeq
Inserting these formulas into Eq. (\ref{s1dh}) we obtain 
1D stationary equations of zero-temperature superfluid hydrodynamics 
\beq
\mu = -{\hbar^2\over 2m \sqrt{n}} \partial_s^2 \sqrt{n} 
+ {m\over 2}(v-v_{\mathrm{B}})^{2} - {\hbar^2\kappa(s)^2\over 8m} 
+ {2a_s(1-1/N) \hbar^2 n \over m l_{\bot}^2} \; , 
\label{q1}
\eeq
and also 
\beq 
\partial_s \left[ n \left( v - v_B \right) \right] = 0 \; . 
\label{q2}
\eeq
Eq. (\ref{q2}) implies that $n \, \left( v - v_B \right) = J$, 
where $J$ is a constant current density. 
This result is very interesting because it says that 
if $n(s)$ has spatial variations then also $v(x)$ must have spatial 
variations. 

Inspired by Ref. \cite{martone}, we now introduce the 
average value of the velocity $v(x)$ in a region $[a,b]$ of the ellipse as 
\beq 
{\bar v} = {1\over (b-a)} \int_{a}^b v(s) \, ds \; ,  
\label{media}
\eeq 
Then, from the previous equations we obtain 
\beq 
{\bar v} = {1\over (b-a)} 
\int_{a}^b \left( {J\over n(s)} + v_B \right) 
\, ds = {J\over {\bar n}_s} + v_B  \; , 
\eeq
where 
\beq 
{\bar n}_{s} = {1\over {1\over (b-a)} \int_a^b {1\over n(s)} ds} \; . 
\label{super-leggett}
\eeq
The number density ${\bar n}_{s}$ can be interpreted 
as the superfluid number density of the stationary state 
in the spatial region $[a,b]$. Indeed, Eq. (\ref{super-leggett}) 
is the 1D version of the formula obtained by Leggett \cite{leggett} 
for a supersolid with spatial periodicity $(b-a)$, 
and recently discussed by others \cite{rica,chomaz,martone}. 
If the stationary state $\Psi(s)$ 
moves with the average velocity ${\bar v}$, its current density reads 
$J = {\bar n}_{s} \left( {\bar v} - v_B\right)$, 
where ${\bar v}$ is the average velocity in the region $[a,b]$
and ${\bar n}_{s}$ the corresponding superfluid number density. 
We can also introduce 
\beq 
{\bar n} = {1\over (b-a)} \int_a^b n(s) \, ds 
\eeq
that is the average number density in the region $[a,b]$. 
Consequently, the superfluid fraction of the BEC in the 
region $[0,L]$ reads
\beq
f_s=\frac{{\bar n}_s}{{\bar n}} = {1\over {N\over L^2}
\int_0^L {1\over n(s)} ds} \; ,  
\label{super-luca}
\eeq
where $N=\int_0^L n(s) \, dx=L{\bar n}$. 
This formula can be also obtained as the response of the 
linear momentum of the BEC to the boost velocity ${\bar v}_B$, 
that is the non-classical translational inertia of the system \cite{rica}. 

\begin{figure}[t]
\begin{center}
\includegraphics[width=12.cm,clip]{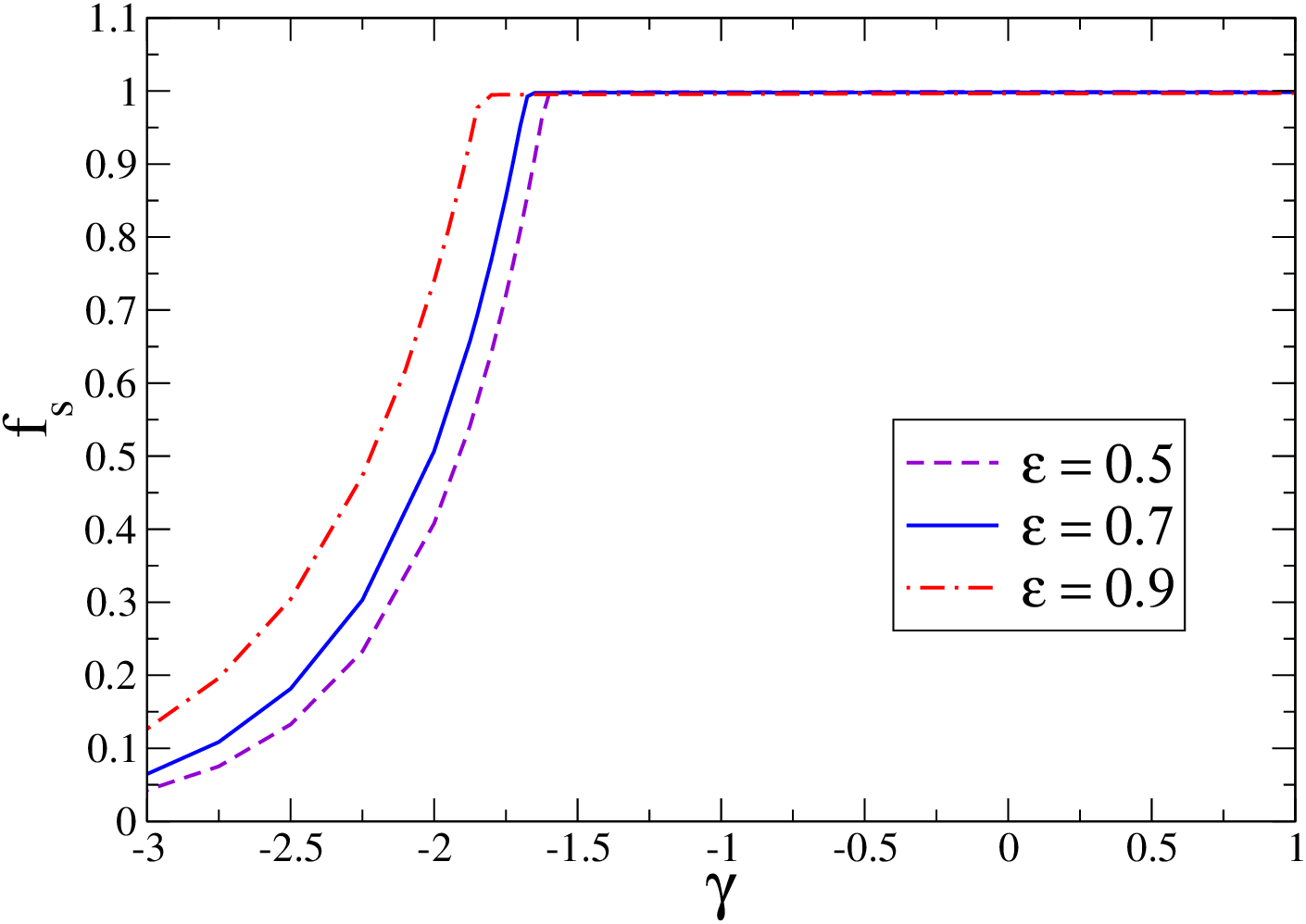}
\end{center}
\caption{Superfluid fraction $f_s$ of the BEC ground state 
in an ellipse as a function of the adimensional interaction 
strength $\gamma=2aa_s(N-1)/l_{\bot}^2$, where $a$ is the length 
of the major semi-axis, $a_s$ is the 3D 
s-wave scattering length and $l_{\bot}$ is the characteristic length of the 
transverse harmonic confinement. The curves are obtained 
for three values of the eccentricity  $\epsilon$ of the ellipse.}
\label{fig4}
\end{figure}

In Fig. \ref{fig4} we plot our numerical results of the 
superfluid fraction $f_s$ as a function of the adimensional 
interaction strength $\gamma$ 
for different values of the eccentricity $\epsilon$ of the ellipse. 
For positive values of $\gamma$ the superfluid fraction $f_s$ is close 
to $1$ also with $\epsilon=0.9$. However, for $0\leq \gamma<1$ and 
a very large eccentricity ($\epsilon=0.99$) we find 
$f_s\simeq 0.95$. This result is quite reasonable because 
the wavefunction is strongly localized in the 
two well of the quantum-curvature potential. For negative values 
of $\gamma$ the most interesting effect appears in Fig. \ref{fig4}: 
around $\gamma \simeq -1.6$ the superfluid fraction $f_s$ 
quckly decreases and it goes to zero for very large negative values 
of $\gamma$. This is exacly the quantum phase transition  
from a two-peak configuration to a one-peak configuration. 
As previously stressed, the one-peak configuration becomes 
modulationally unstable when at least one of the energies of 
its elementary excitations acquires an imaginary component. 
A similar modulational instability \cite{sala-prl}
happens in the formation of a train 
of bright solitons from a single-peak Bose-Einstein 
condensate, induced by a sudden change in the sign of the 
scattering length from positive to negative \cite{nature2002}. 
Our Fig. \ref{fig4} reveals that the critical strength $\gamma_c$ crucially 
depends on the eccentricity $\epsilon$, such as the behavior 
of $f_s$ as a function of $\gamma$ for $\gamma< \gamma_c$. 
In particular, we find that $\gamma_c$ slightly reduces  
by increasing $\epsilon$, but this effect is quite weak.

An important remark is that Eq. (\ref{super-luca}) 
has been derived here without any assumption about 
the sign of $\gamma$. Moreover, the absence of superfluidity 
for $\gamma = 0$ is true only in the thermodynamic limit. In a ring there 
is a finite energy gap between the ground state and the first excited 
state also for $\gamma = 0$. The Bose system we are considering has 
a finite size because it is confined in a finite elliptical ring. 
Finally, the 3D version of Eq. (\ref{super-leggett}) was proposed 
historically by Leggett to characterize the superfluid density  
of a supersolid \cite{leggett}, while here Eq. (\ref{super-leggett}) 
is used to determine the superfluid density of a 
Bose-Einstein condensate which is not supersolid but it is instead 
spatially modulated due to the crucial interplay between elliptical 
confinement and attractive interaction.

\section{Conclusions} 

The main goal of this paper was to understand 
the role of a locally-varying curvature for a Bose-Einstein 
condensate confined in an elliptical waveguide. 
The proposed setup, and the double-well quantum-curvature potential 
that we have found, can be experimentally 
achieved by using ultracold atoms, which are a paradigmatic 
physical platform due to the high experimental tunability of 
inter-atomic interactions and trapping potentials. 
For instance, one can trap $N=10^4$ ultracold Rb atoms by using 
a rapidly moving laser beam which creates a time-averaged 
elliptic-shaped toroidal optical dipole potential \cite{exp1}. The length 
$a$ of the major semi-axis of the ellpse can be $a\simeq 100$ 
microns and the transverse length $l_{\bot}=5$ microns. 
The scattering length $a_s$ could be then tuned by using 
an external magnetic field, which induces a 
Fano-Feshbach resonance \cite{exp2}. 
Despite the fact that we focused on space curvature 
rather than space-time curvature, we believe that our results 
can be of interest not only to atomic and condensed matter 
physics researchers, but also to a large community 
working on general relativity and relativistic quantum field theory. 

{\it Acknowledgements}. The author thanks Francesco Ancilotto, 
Koichiro Furutani, Francesco Minardi, and Andrea Tononi 
for useful discussions.


\begin{thebibliography}{99}


{\cblue 

\bibitem{skinner} E. Switkes, E.L. Russel, and J.L. Skinner, 
Kinetic energy and path curvature in bound state systems,  
J. Chem. Phys. {\bf 67}, 3061 (1977). 

\bibitem{dacosta1} R.C.T. Da Costa, Quantum mechanics of a 
constrained particle, Phys. Rev. A {\bf 23}, 1982 (1981). 

\bibitem{dacosta2} R.C.T. Da Costa, 
Constraints in quantum mechanics. Phys. Rev. A {\bf 25}, 2893 (1982). 

} 

\bibitem{halliwell} J.J. Halliwell, 
Derivation of the Wheeler-DeWitt equation from 
a path integral for minisuperspace models, 
Phys. Rev. D {\bf 38}, 2468 (1988). 

\bibitem{heller} L. Kaplan, N.T. Maitra, and E.J. Heller, 
Quantizing constrained systems, 
Phys. Rev. A {\bf 56}, 2592 (1997). 

\bibitem{lee} J.M. Lee, Riemannian Manifolds: 
An Introduction to the Curvature (Springer, 2007). 

\bibitem{jost} J. Jost, Riemannian Geometry and 
Geometric Analysis (Springer, 2010). 

\bibitem{carmo} M.P. da Carmo, Differential Geometry of 
Curves and Surfaces (Dover Publications, 2016).

{\cblue 

\bibitem{exner1} P. Exner and P. Seba, 
Bound states in curved quantum waveguides, 
J. Math. Phys. {\bf 30}, 2574 (1989).

\bibitem{jaffe} J. Goldstone and R.L. Jaffe, 
Bound states in twisting tubes, Phys. Rev. B {\bf 45}, 14100 (1992). 

\bibitem{clark1} I.J. Clark and A.J. Bracken, 
Effective potentials of quantum strip waveguides and 
their dependence upon torsion, J. Phys. A: Math. Gen. {\bf 29}, 339 (1996). 

\bibitem{clark2} I.J. Clark, 
More on effective potentials of quantum strip waveguides, 
J. Phys. A: Math. Gen. {\bf 31}, 2103 (1998). 

\bibitem{exner2} P. Exner and S.A: Vugalter, 
On the number of particles that a curved quantum waveguide 
can bind, J. Math. Phys. {\bf 40}, 4630 (1999).

\bibitem{exner-book} P. Exner and H. Kovarik, 
Quantum Waveguides (Springer, 2015). 

}

\bibitem{leboeuf} P. Leboeuf and N. Pavloff, 
Bose-Einstein beams: Coherent propagation through a guide, 
Phys. Rev. A {\bf 64}, 033602 (2001). 

\bibitem{schwartz} S. Schwartz, M. Cozzini, C. Menotti, I. Carusotto, 
P. Bouyer, and S. Stringari, 
One-dimensional description of a Bose-Einstein condensate in a 
rotating closed-loop waveguide, 
New J. Phys. {\bf 8}, 162 (2006). 

\bibitem{delcampo} A. del Campo, M.G. Boshier, and A. Saxena, 
Bent waveguides for matter-waves: supersymmetric potentials and 
reflectionless geometries, 
Sci. Rep. {\bf 4}, 5274 (2014). 

\bibitem{sandin} P. Sandin, M. \"Ogren, M. Gulliksson, 
J. Smyrnakis, M. Magiropoulos, and G.M. Kavoulakis, 
Dimensional reduction in Bose-Einstein condensed clouds of atoms confined 
in tight potentials of any geometry and any interaction strength, 
Phys. Rev. E {\bf 95}, 012142 (2017). 

\bibitem{pelster} N.S. Moller, F.E.A. dos Santos, V.S. Bagnato, 
and A. Pelster, 
Bose-Einstein condensation on curved manifolds, 
New. J. Phys. {\bf 22}, 063059 (2020). 
 
\bibitem{sala-prl} L. Salasnich, A. Parola, and L. Reatto, 
Modulational Instability and Complex Dynamics of Confined 
Matter-Wave Solitons, Phys. Rev. Lett. {\bf 91}, 080405 (2003). 

\bibitem{carr} L.D. Carr, C.W. Clark, and W.P. Reinhardt, 
Stationary solutions of the one-dimensional nonlinear Schr\"odinger 
equation. II. Case of attractive nonlinearity, 
Phys. Rev. A {\bf 62}, 063611 (2000).

\bibitem{kav} G.M. Kavoulakis, 
Bose-Einstein condensates with attractive interactions on a ring, 
Phys. Rev. A {\bf 67}, 011601(R) (2003). 

\bibitem{ueda} R. Kanamoto, H. Saito, and M. Ueda, 
Quantum phase transition in one-dimensional Bose-Einstein condensates 
with attractive interactions, 
Phys. Rev. {\bf 67}, 013608 (2003).

\bibitem{leggett} A. J. Leggett, 
Can a Solid Be "Superfluid"?, 
Phys. Rev. Lett. {\bf 25}, 1543 (1970). 

\bibitem{rica} N. Sepulveda, C. Josserand, and S. Rica, 
Nonclassical rotational inertia fraction in a one-dimensional 
model of a supersolid, 
Phys. Rev. B {\bf 77}, 054513 (2008).

\bibitem{chomaz} L. Chomaz, 
Probing the supersolid order via high-energy scattering: 
Analytical relations among the response, density modulation, 
and superfluid fraction, 
Phys. Rev. A {\bf 102}, 023333 (2020). 

\bibitem{martone} G.I. Martone, A. Recati, and N. Pavloff, 
Supersolidity of cnoidal waves in an ultracold Bose gas, 
Phys. Rev. Res. {\bf 3}, 013143 (2021). 

\bibitem{salas} L. Salasnich, A. Parola, and L. Reatto, 
Effective wave-equations for the dynamics of cigar-shaped and disc-shaped 
Bose condensates, 
Phys. Rev. A {\bf 65},  043614 (2002). 

\bibitem{toigo} L. Salasnich, B.A. Malomed, and F. Toigo, 
Matter-wave vortices in cigar-shaped and toroidal waveguides, 
Phys. Rev. A {\bf 76}, 063614 (2007). 

\bibitem{parola} A. Parola, L. Salasnich, R. Rota, and L. Reatto, 
Quantum phases of attractive matter waves in a toroidal trap, 
Phys. Rev. A {\bf 72}, 063612 (2005). 

\bibitem{fattori} A. Trenkwalder, G. Spagnolli, G. Semeghini, 
S. Coop, M. Landini, P. Castilho, L. Pezze, G. Modugno, M. Inguscio, 
A. Smerzi, and M. Fattori, 
Quantum phase transitions with parity-symmetry 
breaking and hysteresis, 
Nature Phys. {\bf 12}, 826 (2016). 
 
\bibitem{nature2002} K.E. Strecker, G.B. Partridge, A.G. Truscott, 
and R.G. Hulet, Formation and propagation of matter-wave soliton trains, 
Nature {\bf 417}, 150 (2002). 

\bibitem{exp1} K. Henderson, C. Ryu, C. MacCormick, and M.G. Boshier, 
Experimental demonstration of painting arbitrary and dynamic potentials 
for Bose–Einstein condensates, 
New J. Phys. {\bf 11}, 043030 (2009). 

\bibitem{exp2} C. Chin, R. Grimm, P. Julienne, and E. Tiesinga, 
Feshbach resonances in ultracold gases, 
Rev. Mod. Phys. {\bf 82}, 1225 (2010). 

\end{thebibliography}
\end{document}